\documentclass[aps,prb,reprint,superscriptaddress]{revtex4-1}

\usepackage[pdftex]{graphicx}
\usepackage{color}
\usepackage{mathtools}

\begin{document}

\title{Static magnetic proximity effects and spin Hall magnetoresistance in Pt/Y$_3$Fe$_5$O$_{12}$ and inverted Y$_3$Fe$_5$O$_{12}$/Pt bilayers}

\author{Stephan Gepr\"{a}gs}
    \email[]{Stephan.Gepraegs@wmi.badw.de}
    \affiliation{Walther-Mei{\ss}ner-Institut, Bayerische Akademie der Wissenschaften, 85748 Garching, Germany}
    \homepage[]{http://www.wmi.badw.de/}
\author{Christoph Klewe}
    \affiliation{Advanced Light Source, Lawrence Berkeley National Laboratory, California 94720, USA}
\author{Sibylle Meyer}
    \affiliation{Walther-Mei{\ss}ner-Institut, Bayerische Akademie der Wissenschaften, 85748 Garching, Germany}		
\author{Dominik Graulich}
    \affiliation{Center for Spinelectronic Materials and Devices, Department of Physics, Bielefeld University, 33615 Bielefeld, Germany}
\author{Felix Schade}
    \affiliation{Walther-Mei{\ss}ner-Institut, Bayerische Akademie der Wissenschaften, 85748 Garching, Germany}
\author{Marc Schneider}
    \affiliation{Walther-Mei{\ss}ner-Institut, Bayerische Akademie der Wissenschaften, 85748 Garching, Germany}
\author{Sonia Francoual}
    \affiliation{Deutsches Elektronen-Synchrotron DESY, 22607 Hamburg, Germany}	
\author{Stephen P. Collins}
    \affiliation{Diamond Light Source Ltd, Harwell Sci \& Innovat Campus, Didcot OX11 0DE, Oxon, England}
\author{Katharina Ollefs}
    \altaffiliation{Present address: Faculty of Physics and Center for Nanointegration Duisburg-Essen (CENIDE), Universit\"{a}t Duisburg-Essen, 47057 Duisburg, Germany}
    \affiliation{European Synchrotron Radiation Facility (ESRF), 38043 Grenoble Cedex 9, France}
\author{Fabrice Wilhelm}
    \affiliation{European Synchrotron Radiation Facility (ESRF), 38043 Grenoble Cedex 9, France}
\author{Andrei Rogalev}
    \affiliation{European Synchrotron Radiation Facility (ESRF), 38043 Grenoble Cedex 9, France}
\author{Yves Joly}
		\affiliation{University Grenoble Alpes, CNRS, Grenoble INP, Institut N\'eel, 38000 Grenoble, France}
\author{Sebastian T.B. Goennenwein}
    \affiliation{Technische Universit\"{a}t Dresden, 01069 Dresden, Germany}
\author{Matthias Opel}
    \email[]{Matthias.Opel@wmi.badw.de}
    \affiliation{Walther-Mei{\ss}ner-Institut, Bayerische Akademie der Wissenschaften, 85748 Garching, Germany}
\author{Timo Kuschel}
    \affiliation{Center for Spinelectronic Materials and Devices, Department of Physics, Bielefeld University, 33615 Bielefeld, Germany}
\author{Rudolf Gross}
    \affiliation{Walther-Mei{\ss}ner-Institut, Bayerische Akademie der Wissenschaften, 85748 Garching, Germany}
    \affiliation{Physik-Department, Technische Universit\"{a}t M\"{u}nchen, 85748 Garching, Germany}
    \affiliation{Munich Center for Quantum Science and Technology (MCQST), 80799 M\"{u}nchen, Germany}

\date{\today}

\begin{abstract}
The magnetic state of heavy metal Pt thin films in proximity to the ferrimagnetic insulator Y$_3$Fe$_5$O$_{12}$ has been investigated systematically by means of x-ray magnetic circular dichroism and x-ray resonant magnetic reflectivity measurements combined with angle-dependent magnetotransport studies. To reveal intermixing effects as the possible cause for induced magnetic moments in Pt, we compare thin film heterostructures with different order of the layer stacking and different interface properties. For standard Pt layers on Y$_3$Fe$_5$O$_{12}$ thin films, we do not detect any static magnetic polarization in Pt. These samples show an angle-dependent magnetoresistance behavior, which is consistent with the established spin Hall magnetoresistance. In contrast, for the inverted layer sequence, Y$_3$Fe$_5$O$_{12}$ thin films grown on Pt layers, Pt displays a finite induced magnetic moment comparable to that of all-metallic Pt/Fe bilayers. This magnetic moment is found to originate from finite intermixing at the Y$_3$Fe$_5$O$_{12}$/Pt interface. As a consequence, we found a complex angle-dependent magnetoresistance indicating a superposition of the spin Hall and the anisotropic magnetoresistance in these type of samples. Both effects can be disentangled from each other due to their different angle dependence and their characteristic temperature evolution.       
\end{abstract}

 %Multilayers
 %Magnetic properties of interfaces (multilayers, superlattices, heterostructures)
 %Spin transport effects

%\keywords{}

\maketitle

\section{Introduction}
\label{sec:intro}

The understanding of spin phenomena in condensed matter is of great importance in the fields of spintronics and spin caloritronics.\cite{Wolf:2001,Bauer:2012} In this context, heavy metal/ferromagnetic insulator (HM/FMI) heterostructures provide a unique platform for the generation and detection of pure spin currents utilizing the (inverse) spin Hall effect (SHE),\cite{Hirsch:1999} spin pumping,\cite{Tserkovnyak:2002} or the spin Seebeck effect.\cite{Uchida:2008} In these HM/FMI heterostructures, the transport of spin angular momentum across the HM/FMI interface is of key importance. Obviously, the magnetic and structural properties at the interface between the HM layer and the FMI play a crucial role for the interfacial exchange interaction between the HM conduction electrons and the localized ions in the FMI.\cite{Tserkovnyak:2005,Puetter:2016,Cramer:2017,Vasili:2018,Zhang:2019,Liu:2020} In this respect, possible static magnetic moments in the HM layer at the interface to the FMI induced by magnetic proximity effects (MPE) came into the focus of research,\cite{Huang:2012, Lu:2013, Geprags:2012,Geprags:2013,Kuschel:2015,Kikkawa:2017} since HMs, such as Pt or Pd, are typically close to the Stoner criterion for ferromagnetism and exhibit proximity induced finite magnetic moments in contact to ferromagnetic metals (FMMs).\cite{Wilhelm:2000,Wilhelm:2001,Wende:2004}  

The MPE can be attributed to a direct exchange interaction between the magnetic elements in the ferromagnet and the conduction electrons of the HM, which is determined by the overlap of their wave functions reflecting the short range nature of the MPE.\cite{Liang:2016, Wende:2004} While the appearance of a finite magnetic polarization in HMs in contact to FMMs is unquestionable,\cite{Wende:2004} the MPE in HM/FMI heterostructures is controversially discussed. In the prototype Pt/Y$_{3}$Fe$_{5}$O$_{12}$ (YIG) structure, Lu and coworkers reported an induced ferromagnetic moment of 0.054~$\mu_{\rm B}$ per Pt atom at room temperature by element-selective x-ray magnetic circular dichroism (XMCD) measurements at the Pt $L_{2,3}$ edges, which even increases to 0.076~$\mu_{\rm B}$ per Pt atom at 20\,K.\cite{Lu:2013} However, the significant ferromagnetic Pt moment at room temperature could not be verified by other groups using Pt or Pd as the HM layer and YIG or ferrites $A$Fe$_2$O$_4$ ($A=$Fe, Ni, Co, Mn) as the FMI.\cite{Geprags:2012,Valvidares:2016,Collet:2017,Kelly:2017,Vasili:2018} In addition, no indication of a finite magnetic polarization in Pt grown on FMIs could be found by x-ray resonant magnetic reflectivity (XRMR) measurements, which is a direct element-specific measure of the spin polarization at the interface. The MPE has been excluded by XRMR for Pt on top of NiFe$_{2}$O$_{4}$ deposited by chemical vapor deposition\cite{Kuschel:2015} and by sputtering.\cite{Kuschel:2016} Recently, however, a field induced magnetic polarization in Pt on YIG was found at low temperature and high magnetic fields by XMCD, mainly caused by the paramagnetic nature of Pt.\cite{Kikkawa:2017} Furthermore, a strong XMCD signal was observed in Fe$_3$O$_4$/Pt/Fe$_3$O$_4$ epitaxial trilayers mainly caused by Fe-Pt interdiffusion and Fe-Pt alloying due to the deposition of Pt at high temperature.\cite{Kikkawa:2019}

The difference between the MPE in HM/FMM and HM/FMI structures seems to be consistent with the situation of superconducting thin films on FMM or FMI: the magnetic moment of Cu in YBa$_2$Cu$_3$O$_7$ is finite when in proximity to the FMM La$_{2/3}$Ca$_{1/3}$MnO$_3$, but below the detection limit on the FMI LaMnO$_3$.\cite{Satapathy:2012}        

The presence or absence of MPEs in HM/FMI heterostructures has direct consequences for the understanding of spin current experiments. Additional magnetoresistance (MR) or magneto-Seebeck and Nernst effects will occur in spintronic and spin caloritronic experiments in the presence of static magnetic moments in the HM layer.\cite{Miao:2016,Bougiatioti:2017} As an example, the MR found in HM/FMI heterostructures was attributed to a magnetic-proximity MR based on the conventional anisotropic magnetoresistance (AMR).\cite{Huang:2012,Lu:2013,Qu:2013} 

Nakayama \textit{et al.}\cite{Nakayama:2013} and Althammer \textit{et al.},\cite{Althammer:2013} however, demonstrated that the longitudinal resistivity of the HM layer reaches its maximum value, if the magnetization of the underlying FMI is either aligned along the current direction $\mathbf{j}$ or the normal $\mathbf{n}$ of the thin film plane, while a minimum value was detected, when aligning the magnetization in the film plane perpendicular to $\mathbf{j}$. This characteristic angle dependence is inconsistent with the conventional AMR of a polycrystalline HM layer\cite{Althammer:2013} and is, instead, explained by an interplay of charge and spin currents at the interface between the FMI and the HM layer via the (inverse) SHE.\cite{Chen:2013} This so-called spin Hall magnetoresistance (SMR) was further experimentally confirmed in a variety of HM/FMI heterostructures such as Pt/YIG,\cite{Nakayama:2013, Althammer:2013, Vlietstra:2013, Hahn:2013, Marmion:2014, Meyer:2014, Aldosary:2016} Ta/YIG,\cite{Hahn:2013}, Pt/Gd$_3$Fe$_5$O$_{12}$,\cite{Ganzhorn:2016} Pt/Fe$_3$O$_4$,\cite{Althammer:2013} Pt/NiFe$_2$O$_4$,\cite{Althammer:2013,Althammer:2019} Pt/CoFe$_2$O$_4$,\cite{Isasa:2014} and Pt/Cu$_2$OSeO$_3$\cite{Aqeel:2016} as well as using antiferromagnetic insulators NiO\cite{Hoogeboom:2017,Fischer:2018,Baldrati:2018,Gepraegs:2020},  Cr$_2$O$_3$,\cite{Ji:2017,Schlitz:2018} and $\alpha$-Fe$_2$O$_3$.\cite{Lebrun:2019, Cheng:2019, Fischer:2020,Gepraegs:2020} The exchange of spin angular momentum as the underlying mechanism of the SMR is further confirmed by Pt/YIG/Pt trilayer structures\cite{Li:2016, Wu:2016} and non-local transport experiments in Pt/YIG bilayer nanostructures.\cite{Cornelissen:2015,Goennenwein:2015,Wimmer:2019}     

As discussed by Kikkawa and coworkers,\cite{Kikkawa:2017} the presence of a finite magnetic polarization in the HM caused by MPEs could be related to defects at the interface between the FMI and the HM layer, such as interdiffusion of ions and alloying, thin amorphous layers, vacancies or free magnetic elements at the surface, demonstrating that the quality of the interface is of crucial importance. This is also confirmed by Vasili \textit{et al.}.\cite{Vasili:2018} They showed that the presence or absence of a MPE in Pt depends on the Pt growth conditions, leading to a possible interfacial reconstruction. 

To get more insight into the origin of MPEs in HM/FMI heterostructures and their correlation to MR effects, we systematically investigate Pt/YIG heterostructures with different order of the layer stacking and different interface properties using element-selective x-ray and angle-resolved magnetotransport studies. We find no indication for any MPE in standard Pt/YIG heterostructures with clean, in-situ grown Pt/YIG interfaces. In these samples, the existing temperature-dependent MR in the Pt layer can be explained within the framework of the SMR model. In contrast, in inverted YIG/Pt heterostructures, we clearly detect a finite XMCD as well as a XRMR signal demonstrating induced magnetic moments in the Pt layer most likely caused by intermixing effects at the YIG/Pt interface. In those samples, we find a superposition of the SMR and the AMR, which can be disentangled from each other due to their characteristic angle and temperature dependence. We further investigated aging effects of the Pt layers utilizing x-ray absorption near edge spectroscopy and magnetotransport measurements. While the white line intensity of the Pt $L_{2,3}$ edges in the inverted YIG/Pt heterostructures stays nearly unaffected over time, a clear increase is observed in standard Pt/YIG bilayers indicating oxidation effects. However, the SMR amplitude changes only marginally with time.      
 
\section{Sample fabrication and characterization}
\label{sec:Preparation-Characterization}

\subsection{Thin film deposition}

A series of Pt/YIG bilayer heterostructures on (111)-oriented, single crystalline Y$_{3}$Al$_{5}$O$_{12}$ (YAG) substrates was fabricated \textit{in-situ} in an ultra-high vacuum system.\cite{Opel:2014} The YIG thin films were deposited by pulsed laser deposition (PLD)\cite{Opel:2012} from a stoichiometric, polycrystalline target using a KrF excimer laser ($\lambda = 248$\,nm) with a fluence of 2\,J/cm$^{2}$ and a repetition rate of 10\,Hz. The Pt layers were deposited via electron-beam evaporation in ultra-high vacuum at room temperature with a deposition rate of $0.4$\,\AA/s. To probe possible intermixing effects at the Pt/YIG interface, we fabricated two different types of bilayer samples: The first one consists of a YIG thin film deposited in O$_2$ atmosphere at $500^\circ$C capped \textit{in-situ} without breaking the vacuum with an approximately 2\,nm thick Pt layer (``standard'' Pt/YIG//YAG bilayer).\cite{Geprags:2012, Althammer:2013} For the second type of bilayers, we first deposited a polycrystalline 10\,nm thin Pt film on a YAG substrate and subsequently, \textit{in-situ}, the YIG thin film on top in an Ar atmosphere at $450^\circ$C to suppress oxidation of the Pt layer (``inverted'' YIG/Pt//YAG bilayer).\cite{Aldosary:2016} 

We expect a clean and sharp interface between the metallic, polycrystalline Pt thin film and the insulating YIG layer for the standard Pt/YIG//YAG bilayer samples,\cite{Geprags:2012} since the electron-beam evaporation method is associated with low, thermal kinetic energies of the Pt particles. Hence, a vanishingly small intermixing at the YIG/Pt interface is expected for this stacking sequence. For the inverted YIG/Pt//YAG bilayer samples, however, the situation is different. Here, Pt is partly incorporated into the YIG thin film and vice versa, due to the high kinetic energies of the atoms and ions in the laser plume during the PLD-process of YIG.\cite{Willmott:2000,Mitra:2017} On the other hand, a partial oxidation of the Pt thin film of the standard Pt/YIG//YAG bilayer samples over time is expected. This is not the case for the inverted YIG/Pt//YAG bilayer samples, since here the thin Pt layer is covered by the thick YIG layer. 

In the following, we will restrict our discussion to the data of two representative standard and inverted bilayer samples.

\subsection{Structural characterization}
\label{sec:structural}

%%
%--------------------------- Fig. 1 --------------------------
\begin{figure}
  \includegraphics[width=\columnwidth]{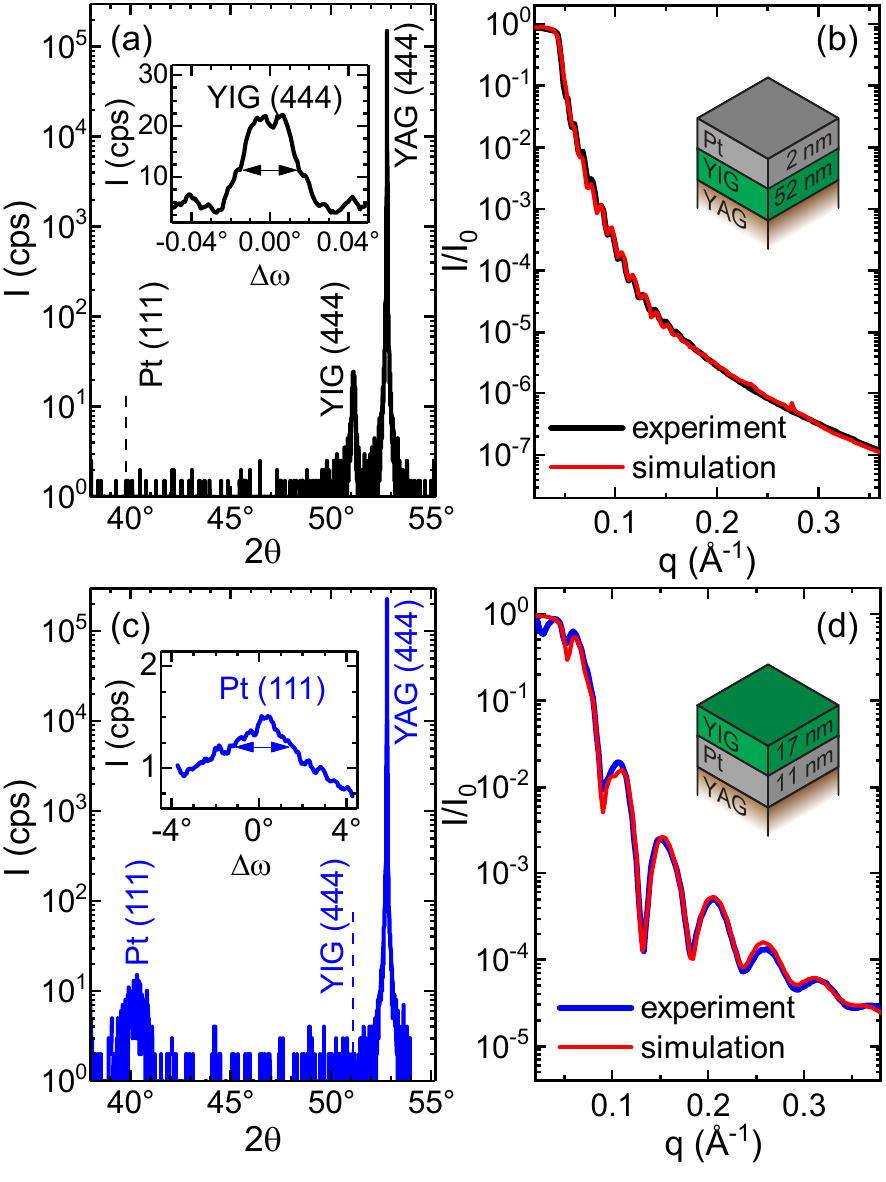}
  \caption{Structural properties of (a),(b) a standard Pt/YIG//YAG bilayer (black lines) and (c),(d) an inverted YIG/Pt//YAG bilayer (blue lines) measured at 300\,K. The out-of-plane high-resolution x-ray $\omega$-$2\theta$ diffraction scans are shown in (a),(c). The expected $2\theta$-positions of the Pt~(111) and YIG~(444) reflections are indicated by dashed vertical lines. The insets display the rocking curves around the YIG~(444) and the Pt~(111) reflections, respectively. (b),(d) Non-magnetic x-ray reflectivity scans plotted against the scattering vector $q$. From the simulations (red lines), we determine the thickness and roughness of the layers. The bilayer stacks with the layer thickness of the respective samples are sketched in the insets.}
	\label{fig:fig1}					
\end{figure}
%%
%--------------------------- Fig. 1 --------------------------

The samples were characterized with respect to their structural properties using high-resolution x-ray diffractometry (HR-XRD) in a four-circle diffractometer with monochromatic Cu $K \alpha_1$ radiation with a wavelength of 0.15406\,nm. The $\omega$-2$\theta$ scans of the bilayers reveal no secondary crystalline phases (cf.~Figs.~\ref{fig:fig1}(a),(c)). However, while YIG was found to be crystalline in the standard Pt/YIG//YAG bilayer with a fully relaxed crystal structure exhibiting a lattice constant of 1.238\,nm,\cite{Geller:1957} no reflection of YIG was found in the inverted YIG/Pt//YAG bilayer indicating a polycrystalline growth of YIG on Pt. For the standard Pt/YIG//YAG bilayer, the rocking curves around the YIG~(444) reflection display a full width at half maximum (FWHM) of about $0.03^\circ$ (cf.~inset in Fig.~\ref{fig:fig1}(a)), indicating a low mosaic spread of the YIG thin films despite the 3\% lattice mismatch between YIG and the YAG substrate. On the other hand, we do not detect any Pt-related reflection, indicating a polycrystalline structure of the Pt thin film. The situation is different for the inverted YIG/Pt//YAG bilayer samples. The $\omega$-2$\theta$ scan displays a weak Pt~(111) reflection, which can be attributed to a (111)-textured Pt layer with a high mosaic spread as revealed by the large FWHM of about $4^\circ$ of the rocking curve (cf.~inset in Fig.~\ref{fig:fig1}(c)). The textured structure arises when heating the Pt thin film to 450$^\circ$C in Ar atmosphere before the deposition of the YIG layer. However, the Pt~(111) reflection is found at a higher $2\theta$-angle ($2\theta=40.21^\circ$) compared to Pt thin films fabricated from the same deposition chamber on various magnetic materials or the literature value of $2\theta = 39.755^\circ$ for bulk Pt.\cite{Arblaster:1997} This might be attributed to a finite intermixing of Pt most likely with Fe rather than oxidation of Pt.\cite{Klemmer:2002, Ellinger:2008}

The thickness as well as an estimation of the roughness of the respective layers were determined by non-magnetic x-ray reflectivity (XRR) at the Diamond Light Source (DLS) at beamline I16 and at the Deutsches Elektronen-Synchrotron (DESY) at beamline P09 with a photon energy of 11566\,eV (cf.~Figs.~\ref{fig:fig1}(b),(d)). From simulations to the experimental data using the recursive Parratt algorithm,\cite{Parratt:1954} a N\'{e}vot-Croce roughness model,\cite{Nevot:1980} and layers with individual refractive indices $n=1-\delta+i\,\beta$ ($\delta$: dispersion, $\beta$: absorption), we obtain $(1.7\pm0.1)$\,nm (Pt) and $(52.0 \pm 0.1)$\,nm (YIG) for the standard Pt/YIG//YAG bilayer (cf.~Fig.~\ref{fig:fig1}(b)). The interface roughness as well as the Pt surface roughness are found to be 0.2\,nm and 1.5\,nm, respectively. Since the Pt and the YIG fringes are not very pronounced, a precise evaluation of the exact structural and optical parameters is challenging resulting in uncertainties of the determined values/significant error bars. Furthermore, as the Pt roughness is in the same range as the total Pt layer thickness, it might influence the resistivity of the Pt layer by surface roughness induced scattering.\cite{Rossnagel:2004} For the inverted YIG/Pt//YAG bilayer sample, we obtain layer thicknesses of YIG and Pt of $(17.4 \pm 0.5)$\,nm and $(11.2 \pm 0.3)$\,nm, respectively (cf.~Fig.~\ref{fig:fig1}(d)). Here, the interface roughness was found to be 0.7\,nm and the YIG surface roughness 1.5\,nm. 

Taken together, the standard Pt/YIG//YAG bilayer samples are composed of an epitaxially grown YIG thin film covered with a polycrystalline Pt thin film, whereas the inverted YIG/Pt//YAG bilayer samples are consisting of a polycrystalline YIG thin film grown on top of a Pt layer exhibiting a (111)-textured structure with possible finite intermixing effects at the YIG interface. 

\subsection{Magnetic characterization}

%%
%--------------------------- Fig. 2 --------------------------
\begin{figure}
  \includegraphics[width=1.0\columnwidth]{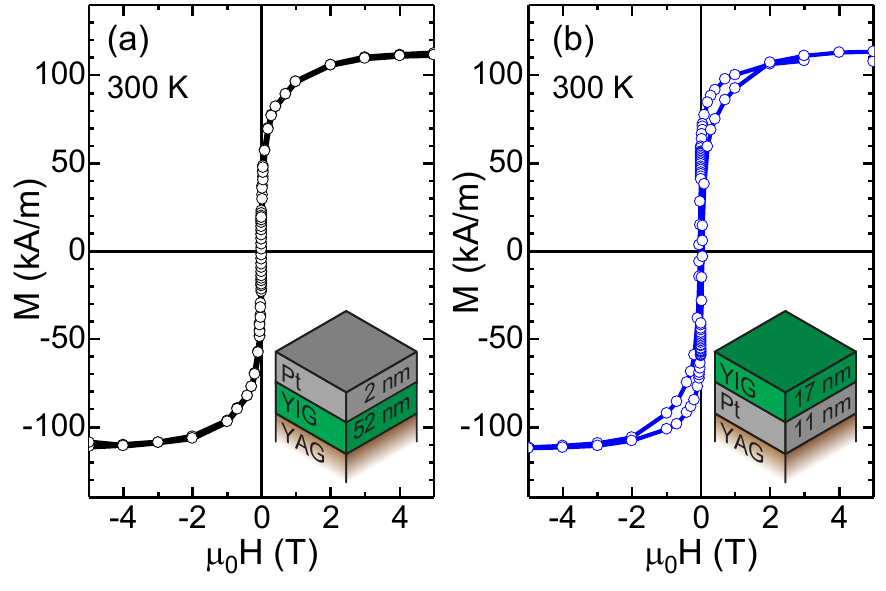}
  \caption{Magnetization curves of (a) a standard Pt/YIG//YAG bilayer (black symbols) and (b) an inverted YIG/Pt//YAG bilayer (blue symbols) recorded at 300\,K with the magnetic field applied parallel to the film plane. The diamagnetic contributions from the YAG substrates have been subtracted.}
	\label{fig:fig2}
\end{figure}
%--------------------------- Fig. 2 --------------------------
%%

The in-plane magnetic properties of the bilayer samples were studied by superconducting quantum interference device (SQUID) magnetometry (cf.~Fig.~\ref{fig:fig2}). At room temperature, we obtain similar magnetization curves for both sample types with a saturation magnetization of around $M_\mathrm{s}=(110\pm5)$\,kA/m. This value is in agreement with our previous results,\cite{Geprags:2012} but slightly lower than the bulk value of $M_{\mathrm{s}}^{\mathrm{YIG}}=143$\,kA/m.\cite{Coey:book} While for the standard Pt/YIG//YAG bilayer sample type, this might be caused by interdiffusion of Al from the YAG substrate into the first monolayers of the YIG thin film,\cite{Popova:2003} the reduced saturation magnetization of the inverted YIG/Pt//YAG sample is most likely also a result of interdiffusion at the YIG/Pt interface. This assumption is supported by the difference in coercive fields of 2\,mT and 40\,mT for the standard Pt/YIG//YAG and the inverted YIG/Pt//YAG sample, respectively. Furthermore, a second magnetically hard phase seems to be present in the inverted sample revealed by the additional hysteresis visible at high magnetic fields between 0.5\,T and 2\,T. This indicates that interdiffusion at the YIG/Pt interface might result in a magnetically hard in-plane component similar to FePt.\cite{Lairson:1993}

\section{Element-selective magnetic properties}
\label{sec:element-selective}

To investigate the magnetic properties of the Pt layers in our bilayer samples, we take advantage of advanced, element-selective, synchrotron-based techniques using hard x-rays.\cite{Ney:2010,Macke:2014} The x-ray absorption near edge spectra (XANES) and XMCD measurements were performed at the European Synchrotron Radiation Facility (ESRF) at the beamline ID12 using the total fluorescence yield detection mode.\cite{Rogalev:2015} The XRMR measurements were carried out at DLS (beamline I16) and DESY (beamline P09).

\subsection{X-ray absorption near edge spectra}
\label{sec:XANES}

%%
%--------------------------- Fig. 3 --------------------------
\begin{figure}
  \includegraphics[width=\columnwidth]{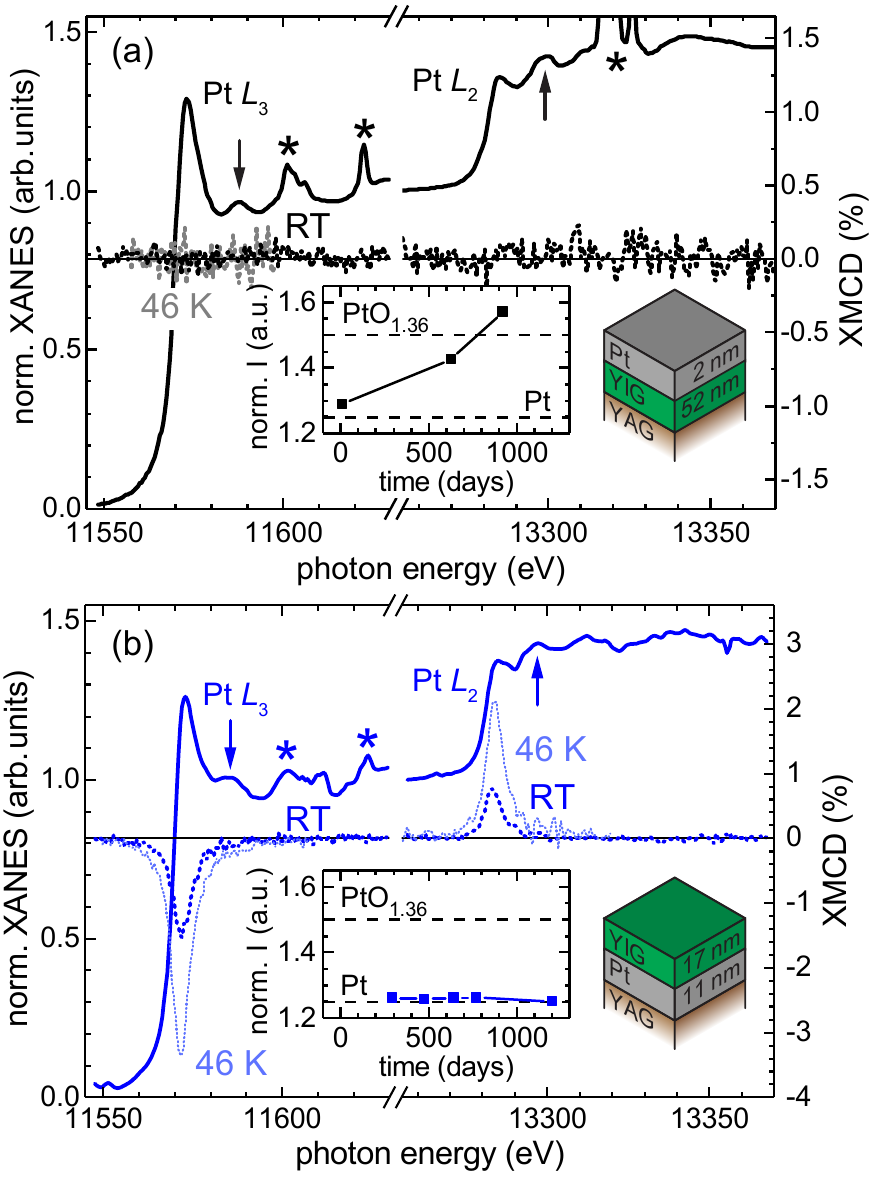}
  \caption{Normalized Pt $L_{2}$ and $L_{3}$ XANES (full lines, left axis) and XMCD spectra (dashed lines, right axis) of (a) a standard Pt/YIG//YAG bilayer (black) and (b) an inverted YIG/Pt//YAG bilayer (blue) measured at room temperature (RT) and 46\,K. The magnetic field of $\pm 0.6$\,T was applied parallel to the incoming x-rays under an angle of 3...5$^\circ$ to the sample surface. Diffraction peaks are marked by asterisks ($\star$). Both samples display EXAFS wiggles at around 11588\,eV and 13300\,eV, marked by the vertical arrows. The inset shows the normalized XANES Pt $L_3$ white line intensity as a function of the time after the sample fabrication. For comparison, the values for Pt and PtO$_{1.36}$ from Ref.~\onlinecite{Kolobov:2005} are indicated by dashed horizontal lines.}
	\label{fig:fig3}
\end{figure}
%--------------------------- Fig. 3 --------------------------
%%
The XANES were recorded around the Pt $L_3$ (11564\,eV) and $L_2$ edges (13273\,eV) with right and left circularly polarized light under positive and negative magnetic fields of $\pm 0.6$\,T and $\pm 0.9$\,T. An electromagnet allowed to flip the field direction at each energy value of the incoming photons. The incident angle of the x-rays was set between $3^\circ$ and $5^\circ$ with respect to the surface plane, while the external magnetic field was aligned parallel to the incident beam. Several XANES were recorded to increase the signal-to-noise ratio and normalized to an $L_3$ edge jump of unity and an $L_2$ edge jump of $0.45$ according to Mattheiss \& Dietz.\cite{Mattheiss:1980}

\begin{figure}
\centering
  \includegraphics[width=\columnwidth]{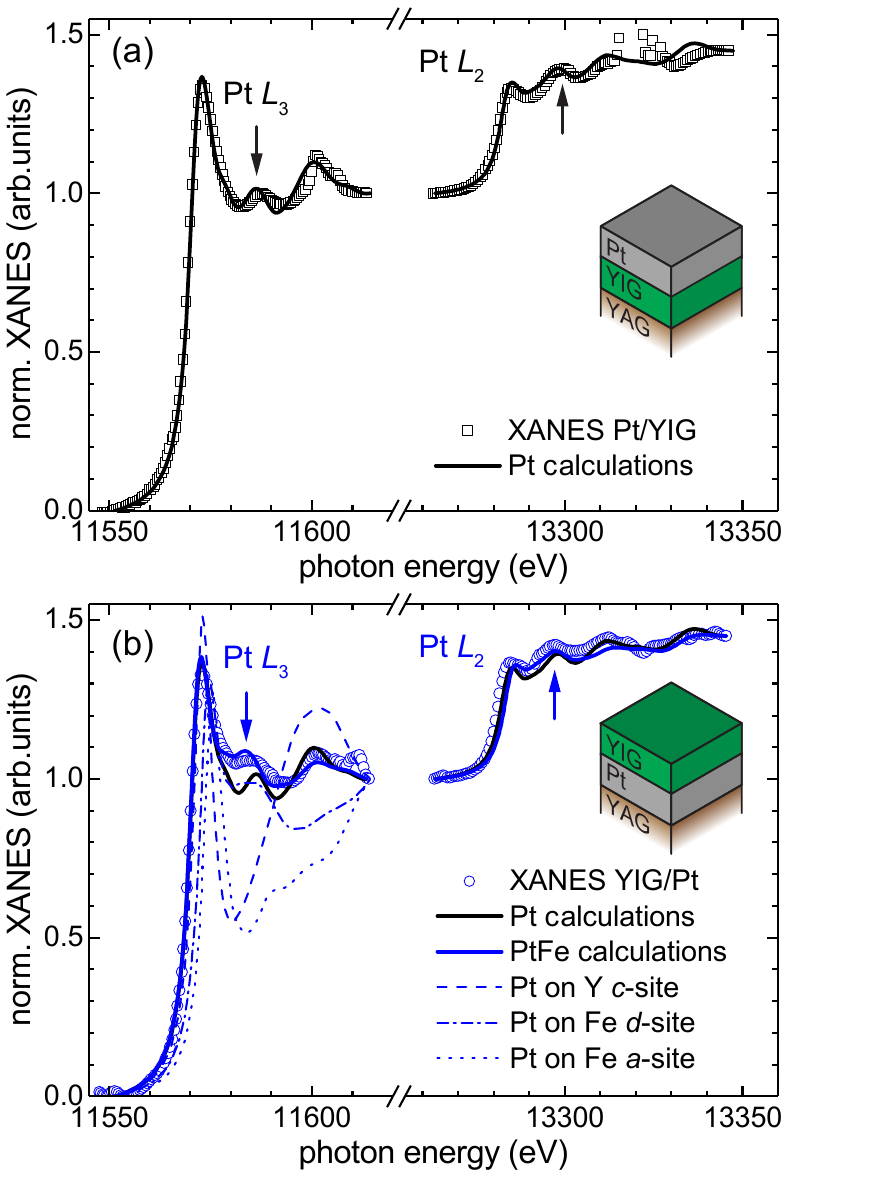}
  \caption{Calculated XANES around the $L_{2}$- and $L_{3}$-edge using the FDMNES code.\cite{FDMNES:webpage} (a) Calculated XANES of Pt (full black line) and experimental XANES (open symbols) of a standard Pt/YIG//YAG bilayer. (b) Calculated XANES of PtFe (full blue line) and Pt-doped YIG with Pt replacing Y on the $c$-site (dashed blue line), substituting Fe on the $d$-site (dashed dotted blue line), as well as on the $a$-site (dotted blue line) of the garnet crystal structure. For comparison, the calculated XANES of Pt (full black line) is also shown. The experimental XANES of an inverted YIG/Pt//YAG bilayer is depicted with open symbols. Since the calculations are in photoelectron energies with zero being the Fermi energy, the calculated XANES were shifted in energy by around 11571\,eV ($L_{3}$-edge) and around 13283\,eV ($L_{2}$-edge), respectively, for comparison with the experimental data. The vertical arrows indicate again the first EXAFS wiggles.}
	\label{fig:fig4}
\end{figure}

The XANES from a standard Pt/YIG//YAG (solid black line) and an inverted YIG/Pt//YAG bilayer (solid blue line) recorded shortly after the fabrication of the samples are shown in Figs.~\ref{fig:fig3}(a) and (b), respectively. We clearly observe the Pt $L_3$ and Pt $L_2$ absorption edges for both sample types. Both XANES also display extended x-ray absorption fine structure (EXAFS) wiggles at around 11588\,eV and 13300\,eV (cf.~vertical arrows in Figs.~\ref{fig:fig3}(a) and (b)). However, these EXAFS wiggles are shifted to lower energies in case of the inverted YIG/Pt//YAG bilayer sample. To obtain more information about the valence state and chemical environment of the absorbing Pt atoms, we calculated the XANES using the FDMNES code\cite{Bunua:2009,FDMNES:webpage} in a Density Functional Theory (DFT) full potential, relativistic approach including spin-orbit coupling (Fig.~\ref{fig:fig4}). This code is extensively used to simulate XANES and resonant x-ray scattering spectra. As obvious from Fig.~\ref{fig:fig4}(a), the calculated XANES of Pt (full black line) is in good agreement with the experimentally obtained XANES of our standard Pt/YIG//YAG bilayer (open black symbols). Not only the Pt $L_3$ and Pt $L_2$ absorption edges but also the EXAFS wiggles are clearly reproduced. In case of the inverted YIG/Pt//YAG bilayer (see~Fig.~\ref{fig:fig4}(b)), the measured XANES (open blue symbols) show clear differences to the simulated Pt XANES (black full line) in the energy range between 11576\,eV and 11595\,eV ($L_3$) as well as between 13285\,eV and 13304\,eV ($L_2$), i.e.~around the first EXAFS wiggle directly after the $L_3$ and $L_2$ absorption edges (cf.~Fig.~\ref{fig:fig4}(b)). In particular, an increased intensity of the EXAFS wiggles at around 11588\,eV and 13300\,eV can be observed and furthermore the EXAFS wiggles are shifted to lower energies. To explain this, we considered a finite interdiffusion at the YIG/Pt interface either leading to regions with a PtFe alloy or a finite Pt-doping of YIG. In the latter case, the calculated XANES of Pt on the different sites of the YIG crystal structure reveal completely different energy dependencies of the x-ray absorbed intensity (cf.~dashed, dashed-dotted, dotted lines in Fig.~\ref{fig:fig4}(b)). However, the calculated XANES of PtFe (blue full line in Fig.~\ref{fig:fig4}(b)) agrees fairly well with the experimental results of the inverted YIG/Pt//YAG bilayer reproducing the higher intensity as well as the energy shift of the EXAFS wiggles. This is in agreement with experiments on Pt alloys\cite{Mukerjee:1995} and indicates that a finite interdiffusion at the YIG/Pt interface of the inverted YIG/Pt//YAG bilayer took place during the deposition leading to regions, where Pt is in direct contact to Fe.  

The white line intensities at the absorption edges represent another known sensitive measure for the valence state as well as chemical environment of Pt, since it is related to the $5d$ electron vacancies.\cite{Horsley:1982} For example, PtO$_{1.6}$ displays an $L_3$ white line intensity of 2.20, PtO$_{1.36}$ of 1.50, and metallic Pt of 1.25, with respect to the edge jump.\cite{Kolobov:2005} Furthermore, a small reduction of the white line intensity to around 1.20 is found for FePt, which is attributed to electronic hybrid states as a result of the alloying of Pt with Fe.\cite{Feng-Ju:2010} We observe values of 1.29 and 1.26 in our standard and inverted bilayer sample types, respectively. These values are very close to our previous observations \cite{Geprags:2012} and the one reported for metallic Pt.\cite{Kolobov:2005} For Pt on differently prepared NiFe$_2$O$_4$ films, slightly larger white line intensities between 1.33 and 1.35 have been found,\cite{Kuschel:2015,Kuschel:2016} while for Pt/CoFe$_2$O$_4$ the white line intensity was 1.24.\cite{Valvidares:2016} However, Lu \textit{et al.} reported a larger white line intensity of 1.45 relative to the edge jump in a 1.5\,nm thin Pt layer on YIG.\cite{Lu:2013} This discrepancy compared to our results and to the literature values is even more pronounced for the white line intensity at the Pt $L_2$ edge. For Pt metal an intensity of $0.79$\cite{Bartolome:2009} is expected, which is in nice agreement with our results (cf.~Fig.~\ref{fig:fig3}(a)), but in contrast to the data reported by Lu \textit{et al.}. They found a value larger than 1.0.\cite{Lu:2013} One possible reason for an enhanced white line intensity might be aging of the Pt surface. For the Pt $L_3$ white lines, we observe an increase in intensity to 1.57 within $\sim 900$\,days after the thin film deposition for the standard bilayer sample type, whereas the value for the inverted one stays constant over time (cf.~insets in Fig.~\ref{fig:fig3}(a),(b)). This can be understood as partial oxidation of Pt when exposed to ambient atmosphere, which is highly possible for the standard Pt/YIG//YAG bilayer, but suppressed by the YIG capping layer in the inverted YIG/Pt//YAG bilayer. Recently, possible aging effects have also been reported for Pt(3.2\,nm)/Fe(9.1\,nm) bilayers, where the magnitude of the induced magnetic Pt moment decreased by about 30\% within half a year.\cite{Kuschel:2015,Klewe:2016}

%%
%--------------------------- Fig. 4 --------------------------
%--------------------------- Fig. 4 --------------------------
%%

\subsection{X-ray magnetic circular dichroism}
\label{sec:XMCD}

The XMCD spectra were calculated as the direct difference between consecutive XANES recorded with opposite x-ray helicity or magnetic field direction. The following XMCD results, therefore, give access only to the projection of the magnetization of Pt on the external magnetic field, i.e. the $k$-vector of the incoming x-ray beam. 

While the XANES of the Pt layer of both sample types are almost identical, their XMCD spectra are different (cf.~dashed lines in Figs.~\ref{fig:fig3}(a) and (b)). For the standard Pt/YIG//YAG bilayer, we do not observe a finite XMCD signal at both Pt $L_{2}$ and $L_{3}$ edges down to a noise level of $<\,0.1\%$ with respect to the edge jump at room temperature (RT) (cf.~black dashed line in Fig.~\ref{fig:fig3}(a)) and 46\,K (cf.~grey dotted line in Fig.~\ref{fig:fig3}(a)). Therefore, we do not find any indication for a finite induced magnetic moment in Pt on YIG due to MPEs. This supports our previous results from a comprehensive XMCD study of three different standard Pt/YIG bilayer samples with different thicknesses (3\,nm, 7\,nm, and 10\,nm) of the Pt top layer, from which we identified an upper limit of a possible induced moment of $(0.003 \pm 0.001)\,\mu_{\rm B}$ per Pt.\cite{Geprags:2012} Furthermore, it is in agreement with the data reported recently by different groups.\cite{Valvidares:2016,Collet:2017,Kelly:2017,Vasili:2018} However, these results are in contrast to the finite induced ferromagnetic magnetic moment of 0.054\,$\mu_\mathrm{B}$ in Pt detected by XMCD in Pt/YIG bilayers reported by Lu and coworkers.\cite{Lu:2013} We note that the noise level of our data is at least 10 times lower than their XMCD signal of $1\%$.

In the inverted YIG/Pt//YAG bilayer sample, however, we detect a finite XMCD signal at both Pt $L_3$ and $L_2$ edges (cf.~Fig.~\ref{fig:fig3}(b)). The maxima of the XMCD signal are located at slightly lower energies than the maximum of the XANES in accordance with literature.\cite{Schuetz:1990} To quantify the induced magnetic moment of the Pt atoms averaged over the Pt film thickness, we apply magneto-optic sum rules neglecting the magnetic dipole term due to the cubic symmetry and the polycrystalline nature of the Pt thin film.\cite{Thole:1992,Carra:1993} To this end, we first determine the number of holes following the method proposed by Ref.~\onlinecite{Grange:1998} using the Au x-ray absorption white line intensity published in Ref.~\onlinecite{Bartolome:2009} as reference. We find a number of holes per Pt atom of 1.87, which is slightly larger than reported for metallic Pt.\cite{Bartolome:2009} With this value, we obtain a spin moment of $m_s = 0.058\,\mu_\mathrm{B}$/Pt and an orbital magnetic moment of $m_l = 0.011\,\mu_\mathrm{B}$/Pt, resulting in a ratio of $m_l/m_s = 0.188$ at room temperature. Thus, $m_s$ is of the same order as the total induced magnetic moment of $0.032\,\mu_\mathrm{B}$/Pt obtained earlier from an all-metallic polycrystalline Pt/Fe reference sample with a 10\,nm thin Pt top layer.\cite{Geprags:2012} This high $m_s$ value suggests that a large fraction of Pt atoms is in direct proximity to Fe atoms, indicating a high level of intermixing at the YIG/Pt interface in the inverted YIG/Pt//YAG bilayer sample. The XMCD signal even increases at lower temperature (cf.~dotted blue line in Fig.~\ref{fig:fig3}(b)). At 46\,K, we find a spin and angular moment of $m_s = 0.160\,\mu_\mathrm{B}$/Pt and $m_l = 0.016\,\mu_\mathrm{B}$/Pt, respectively, which is almost three times larger than at room temperature.

\subsection{X-ray resonant magnetic reflectivity}
\label{sec:XRMR}

We further confirm the XMCD results by XRMR measurements recorded at room temperature at the Pt $L_3$ edge at a fixed photon energy close below the XANES maximum.\cite{Kuschel:2015, Klewe:2016, Kuschel:2016} An external magnetic field of $\mu_0 H=85$\,mT was applied in the scattering plane parallel to the sample surface by a four-coil electromagnet. Circularly polarized light was used for the measurements with a degree of polarization of (81$\pm$5)\% at DLS (standard Pt/YIG//YAG sample) and of (99$\pm$1)\% at DESY (inverted YIG/Pt//YAG bilayer sample). The magnetic field direction was flipped ($\pm\mathbf{H}$) at DLS for each value of the scattering vector $q$, while the reflected intensity $I_{\pm}$ was detected and the x-ray polarization (left) was kept constant. At DESY, the magnetic field stayed constant ($+\mathbf{H}$), while the x-ray helicity was flipped (left/right) for each value of $q$, thus obtaining the reflected intensity $I_{\pm}$.

The XRMR asymmetry ratio $\Delta I=(I_+-I_-)/(I_++I_-)$ is simulated with ReMagX\cite{Macke:2014} using the structural parameters from the non-magnetic XRR analysis (cf.~section~\ref{sec:structural}) and additional magneto-optic depth profiles, which describe the magneto-optic parameters $\Delta \delta$ and $\Delta \beta$ vertical to the layer stack. Slightly different energy calibrations of the beamlines lead to slight variations of the whiteline energies. Therefore, different $\Delta \beta/\Delta \delta$ ratios for analyzing the XRMR data have been chosen by either comparing the XAS to the ab initio calculations used in Ref.~\onlinecite{Kuschel:2015} or by identifying the energy with maximal dichroic effect and vanishing $\Delta \delta$. We use a $\Delta \beta/\Delta \delta$ ratio of 7.3 for fitting the DLS data and $\Delta \delta = 0$ for the DESY data (cf. Ref.~\onlinecite{Bougiatioti:2018}). A detailed description of the fitting procedure and more information on the shape of the magneto-optic depth profiles can be found in Ref.~\onlinecite{Klewe:2016}. Note that $\Delta \delta$ and $\Delta \beta$ are proportional to the magnetic Pt moment and can be converted from magneto-optic constant into a magnetic moment per Pt atom using the conversion factor theoretically calculated in Ref.~\onlinecite{Kuschel:2015} for smooth standard Pt/FMM bilayers as confirmed by XMCD.\cite{Graulich:2020} However, this factor has to be handled with care when analyzing the inverted YIG/Pt//YAG sample with probable interdiffusion at the YIG-Pt interface. Therefore, the focus of the XRMR analysis for the inverted YIG/Pt//YAG sample is the absence or presence of the MPE rather than a rigorous quantitative evaluation. 

%%
%--------------------------- Fig. 5 --------------------------
\begin{figure}
  \includegraphics[width=1.0\columnwidth]{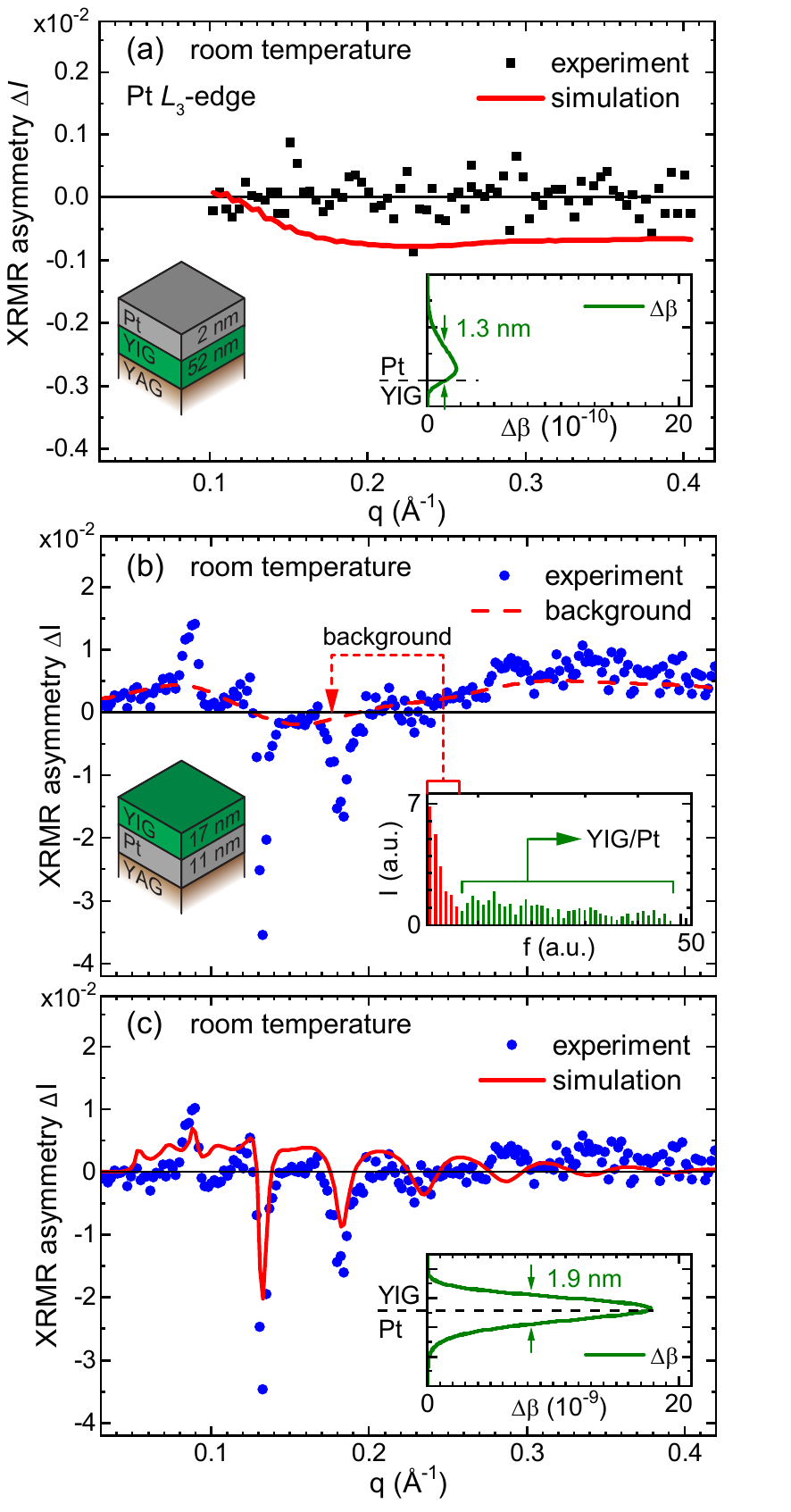}
  \caption{XRMR asymmetry ratios $\Delta I$ at the Pt $L_3$ edge of (a) a standard Pt/YIG//YAG bilayer (black symbols) and (b),(c) an inverted YIG/Pt//YAG bilayer (blue symbols) between XRR curves measured under the same conditions as in Fig.~\ref{fig:fig1}(b),(d), but with an applied magnetic field of $\pm 85$\,mT at room temperature. (a) The asymmetry ratio of the standard bilayer does not show any XRMR response within experimental error. The simulation (red line) using the magneto-optic depth profile (inset) gives an upper limit of 0.002\,$\mu_\mathrm{B}$ per spin-polarized Pt atom. (b) Raw data of the inverted bilayer shows clear oscillations in the asymmetry ratio $\Delta I$. A Fourier analysis (inset) identifies two main contributions: oscillations from the YIG/Pt (green) and a background at very low frequencies (red), which was extracted and back-transformed (dashed red line). (c) Asymmetry ratio after subtraction of the background contribution from (b). The simulation (red line) is based on the magneto-optic depth profile shown in the inset.}
	\label{fig:fig5}
\end{figure}
%--------------------------- Fig. 5 --------------------------
%%

The XRMR responses of the two investigated sample types are quite different as presented in Fig.~\ref{fig:fig5}. The experimental asymmetry ratio $\Delta I$ of the standard sample type (black symbols in Fig.~\ref{fig:fig5}(a)) does not exhibit any oscillations. In comparison, we simulate a theoretical asymmetry ratio of a Pt/YIG//YAG sample with the same structural and optical parameters, but using the magneto-optic depth profile shown in the inset of Fig.~\ref{fig:fig5}(a). The 1.3\,nm effective layer thickness of the spin-polarized Pt is chosen according to the spin-polarized effective layer thickness found for Pt/Fe bilayers.\cite{Kuschel:2015, Klewe:2016, Kuschel:2016} Our simulation yields a long-range oscillation (red line in Fig.~\ref{fig:fig5}(a)) caused by the small Pt thickness of 1.7\,nm. The experimental data clearly do not show such an oscillation or any other non-zero oscillating feature aside from noise fluctuations. This is in accordance with XRMR measurements on Pt/NiFe$_2$O$_4$\cite{Kuschel:2015, Kuschel:2016} and confirms the XMCD results discussed in section~\ref{sec:XMCD}. 

Using the simulated maximum magneto-optic change in Pt as well as the conversion factor determined in Ref.~\onlinecite{Kuschel:2015} and taking into account the degree of light polarization, we calculate an upper limit of 0.002\,$\mu_\mathrm{B}$ per spin-polarized Pt atom within the 1.3\,nm thick spin-polarized volume. If the whole Pt film of 1.7\,nm (spin-polarized plus non-polarized part) was taken into account, the upper limit would be $0.002\,\mu_\mathrm{B}\times$(1.3\,nm/1.7\,nm) $=0.0015\,\mu_\mathrm{B}$ per Pt atom, which improves our sensitivity limit by a factor of two with respect to the XMCD results. This advantage of XRMR is even more pronounced for thicker Pt films, since XRMR is directly sensitive to the spin-polarized Pt, while XMCD is affected by both spin-polarized and non-polarized Pt, as discussed in Refs.~\onlinecite{Kuschel:2015, Klewe:2016}.

In contrast to the standard sample type, the XRMR asymmetry ratio $\Delta I$ of the inverted YIG/Pt//YAG sample (cf.~blue symbols in Fig.~\ref{fig:fig5}(b),(c)) shows clear oscillations due to magnetic Pt. The absolute asymmetry ratio reaches values of up to 3.5\%, which is the same order of magnitude as found for Pt on ferromagnetic metals such as Pt/Fe\cite{Kuschel:2015, Kuschel:2016} or Pt/Ni$_{33}$Fe$_{67}$ and Pt/Ni$_{81}$Fe$_{19}$ (permalloy).\cite{Klewe:2016} By fitting the raw XRMR asymmetry we are able to reproduce the high frequency oscillations from the magnetized Pt. However, we observe a systematic deviation between the simulation and the asymmetry ratio. A Fourier analysis of the XRMR asymmetry (inset of Fig.~\ref{fig:fig5}(b)) reveals an additional low frequency background (red) on top of the magnetic response from YIG/Pt (green) that cannot be reproduced within our model. The background was separated from the other frequencies and back-transformed (cf.~red dashed line in Fig.~\ref{fig:fig5}(b)). This long-range background, which was not observed in previous measurements, will be investigated in future studies.

In order to model the XRMR asymmetry ratio based on the structural and optical parameters of the XRR results, the long-range background is subtracted as shown in Fig.~\ref{fig:fig5}(c). Using the magneto-optical depth profile given in the inset of Fig.~\ref{fig:fig5}(c), we obtain a spin-polarized Pt layer with an effective thickness of 1.9\,nm and a magnetic moment of 0.09\,$\mu_\mathrm{B}$ per spin-polarized Pt atom, which is located near the YIG-Pt interface. Taking the whole Pt thickness of 11.2\,nm into account, we obtain 0.09\,$\mu_\mathrm{B} \times$ 1.9\,nm / 11.2\,nm = 0.015\,$\mu_\mathrm{B}$ per Pt atom, which is in the same order of magnitude as the obtained magnetic moment by the XMCD experiment.

The XRMR results of the inverted YIG/Pt//YAG bilayer are not as comparable to XMCD as the results that have been obtained for Pt grown on Fe and CoFe with almost perfect quantitative agreement between XRMR and XMCD ($\pm$2\%).\cite{Graulich:2020} One reason could be the poorer fit quality of the asymmetry ratio for the inverted YIG/Pt//YAG bilayer compared to metallic bilayers. Still, the fit result presented here reflects a clear overall global minimum as tested by goodness-of-fit space mappings which have proven effective in previous XRMR studies.\cite{Krieft:2020,Moskaltsova:2020} In addition, the origin of the oscillating background observed for the inverted YIG/Pt//YAG bilayer is unknown and has not been observed previously for metallic bilayers, but could affect the quantitative XRMR results as well. Moreover, the high roughness of the YIG-Pt interface together with the possible interdiffusion at this interface is not captured by the ab initio calculations that form the basis of the $\Delta \beta$ to $\mu_\mathrm{B}$ per Pt atom conversion factor.\cite{Kuschel:2015} In spite of these uncertainties and the discrepancy between the quantitative XRMR and XMCD results, we clearly observe spin-polarized Pt at the YIG-Pt interface of the inverted YIG/Pt//YAG sample.

\section{Magnetotransport measurements}

%%
%--------------------------- Fig. 6 --------------------------
\begin{figure*}[t]
  \includegraphics[width=17cm]{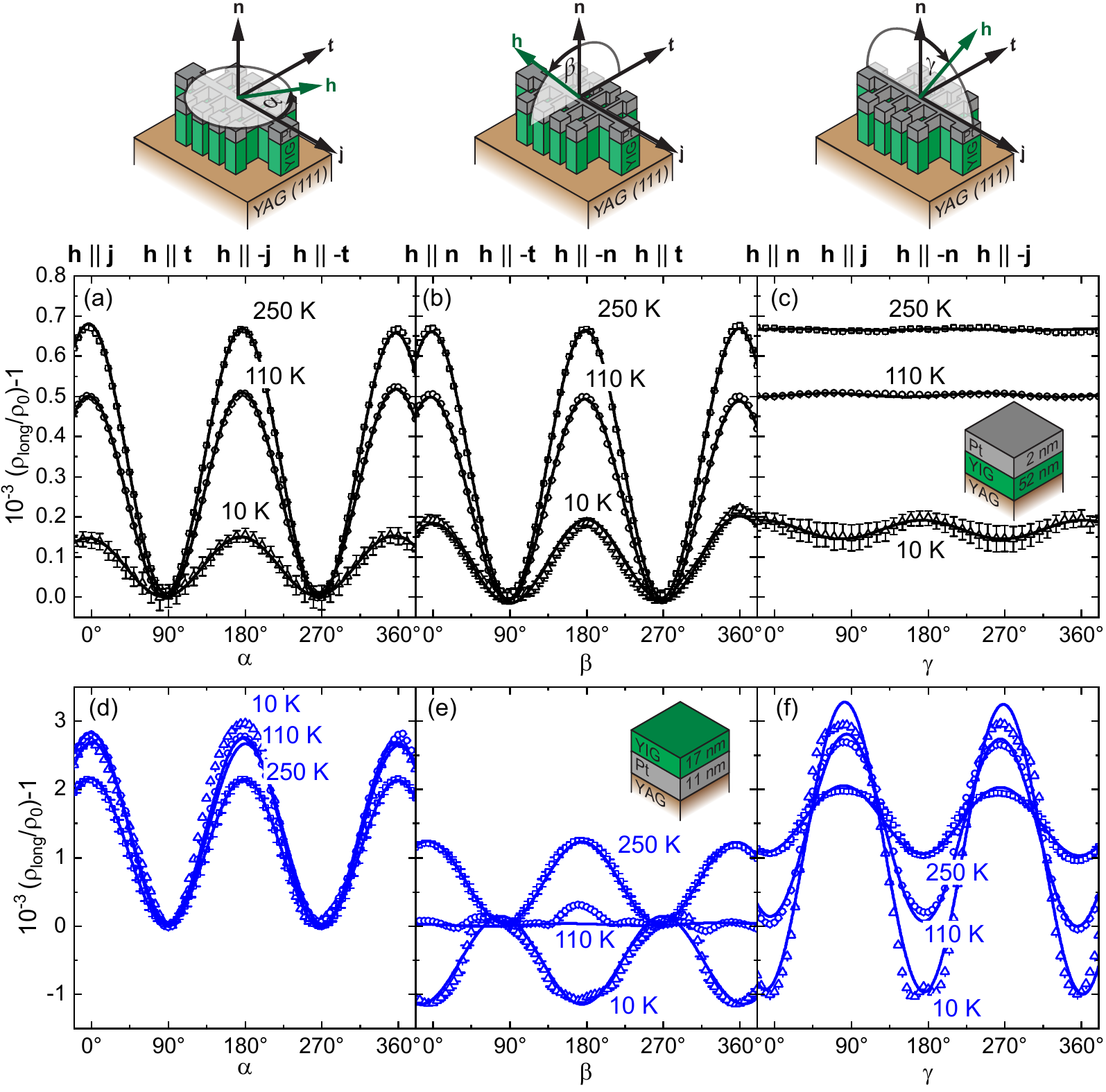}
  \caption{Normalized longitudinal resistivity $\rho_\mathrm{long}$ of (a),(b),(c) a standard Pt/YIG//YAG bilayer sample (black symbols) and (d),(e),(f) an inverted YIG/Pt//YAG bilayer sample (blue symbols) recorded at 250\,K, 110\,K, and 10\,K in an external magnetic field of $\mu_0H=7$\,T. The magnetic field $\mathbf{H}$ is rotated (a),(d) in-plane (ip-rotation), (b),(e) out-of-plane perpendicular to the current direction $\mathbf{j}$ (oopj-rotation), and (c),(f) out-of-plane perpendicular to the transverse direction $\mathbf{t}$ (oopt-rotation), as illustrated above. The lines are fits to the data using $\cos^2$-functions (cf.~Eqs.~(\ref{eq:SMR}) and (\ref{eq:AMR})). All curves are normalized to $\rho_0=\rho_\mathrm{long, H\|t}$.}
 \label{fig:fig6}
\end{figure*}
%--------------------------- Fig. 6 --------------------------
%%

To investigate the impact of the finite induced magnetic Pt-moment of the inverted YIG/Pt//YAG bilayer sample on the SMR, the thin film bilayer samples were patterned into Hall bar mesa structures (width $\mathrm{80\,\mu m}$ and contact separation $\mathrm{800\,\mu m}$) via photolithography and Ar ion milling. The samples were mounted on a rotatable sample holder, which is placed in a superconducting magnet cryostat. We applied a current $I_c=100\;\mu$A in the Pt layer with the current direction $\mathbf{j} \parallel [1\overline{2}1]$ of the YIG thin film and measured the longitudinal voltage $V_\mathrm{long}$ via a DC current reversal technique,\cite{Ganzhorn:2016} while rotating the magnetic field $\mathbf{H}$ at a constant magnitude of $\mu_0H = 7$\,T. This value is well above the saturation field of YIG (cf.~Fig.~\ref{fig:fig2}), ensuring that the magnetization $\mathbf{M}$ of the YIG layer is saturated along the magnetic field direction $\mathbf{H}$. We rotated the magnetic field $\mathbf{H}$ in three different rotation planes: (i) in the film plane with angle $\alpha$  (ip), (ii) out of the film plane perpendicular to the current direction $\mathbf{j}$ with angle $\beta$ (oopj), and (iii) out of the film plane perpendicular to the transverse direction $\mathbf{t}$ with angle $\gamma$ (oopt) (cf.~insets in Fig.~\ref{fig:fig6}(a)-(c)). From the measured longitudinal voltage, we calculated the longitudinal resistivity $\rho_\mathrm{long}$ and normalized it to the respective value $\rho_0=\rho_\mathrm{long, H\|t}$. $\rho_0$ differs significantly for both sample types. While for the inverted YIG/Pt//YAG bilayer $\rho = 2.71 \times 10^{-7}\,\mathrm{\Omega m}$ at 300\,K is in perfect agreement with previous reports,\cite{Meyer:2014} an increased resistivity of $\rho = 1.19 \times 10^{-6}\,\mathrm{\Omega m}$ at 300\,K is found for the standard Pt/YIG//YAG bilayer sample type. This can be attributed to size as well as roughness effects, and can be described in a modified Fuchs-Sondheimer theory.\cite{Fuchs:1938,Sondheimer:1952,Fischer:1980} Moreover, the higher resistivity of the standard bilayer sample type further indicates a partially oxidized Pt layer. 

The angle-dependent magnetoresistance (ADMR) of the standard Pt/YIG//YAG as well as the inverted YIG/Pt//YAG bilayer sample types carried out in the three rotation planes (ip, oopj, oopt) are shown in Fig.~\ref{fig:fig6}. The ADMR of the standard bilayer is consistent with the SMR model as discussed in detail in Refs.~\onlinecite{Althammer:2013,Chen:2013}. Within this model, the modulation of $\rho_\mathrm{long}$ as a function of the magnetization direction $\mathbf{m}=\mathbf{M}/M_s$ of the YIG layer reflects the interplay between charge and spin currents at the interface between Pt and YIG via the (inverse) SHE. A charge current density $\mathbf{J}_c$ in the conducting Pt layer induces a spin current density $\mathbf{J}_s$ perpendicular to the spin polarization $\boldsymbol{\sigma}$ and $\mathbf{J}_c$ via the SHE. This results in a local spin accumulation at the Pt/YIG interface if $\boldsymbol{\sigma}$ is collinear to $\mathbf{m}$, which induces a diffusive spin current backflow $\mathbf{J}_s^\mathrm{back}$ compensating $\mathbf{J}_s$. If $\boldsymbol{\sigma}$ is non-collinear to $\mathbf{m}$, a spin transfer torque is exerted on the magnetic moments, reducing the spin accumulation. This results in an additional dissipation channel for charge transport in the Pt layer leading to an increase of the Pt resistivity.\cite{Nakayama:2013,Althammer:2013} The modulation of the component of the Pt resistivity tensor $\boldsymbol{\rho}$ along the current direction $\mathbf{j}$, coinciding with the longitudinal resistivity $\rho _{\mathrm{long}}$, is given by\cite{Chen:2013} 
\begin{eqnarray}
  \rho^\mathrm{SMR}_\mathrm{long}&=&\rho^\mathrm{SMR}_{0}+ \rho^\mathrm{SMR}_{1}\left[1 - m_{t}^2\right]  
	\label{eq:SMR}
	\; ,
\end{eqnarray}       
where $\rho^\mathrm{SMR}_{0}$ is approximately equal to the normal resistivity of the Pt layer.\cite{Chen:2013} $\rho^\mathrm{SMR}_{1}$ represents the SMR coefficient with $\rho^\mathrm{SMR}_{1} \ll \rho^\mathrm{SMR}_{0}$, and $m_{t}$ denotes the projection of $\mathbf{m}$ on $\mathbf{t}$. Here we assume that the SMR amplitudes $\rho^\mathrm{SMR}_{1,1}$ and $\rho^\mathrm{SMR}_{1,2}$ of the two strongly antiferromagnetically coupled Fe sublattices in the collinear ferrimagnetic state of YIG are equal: $\rho^\mathrm{SMR}_{1}=\rho^\mathrm{SMR}_{1,1}=\rho^\mathrm{SMR}_{1,2}$.\cite{Ganzhorn:2016,Fischer:2018,Gepraegs:2020} 

Equation~\ref{eq:SMR} results in a $\cos^2\alpha$ and $\cos^2\beta$ dependence of $\rho_\mathrm{long}$, when rotating the magnetic field $\mathbf{H}$ in the ip- and oopj-plane, while no ADMR of $\rho_\mathrm{long}$ is expected for magnetic field rotations in the oopt-plane.\cite{Althammer:2013} This is fundamentally different to the anisotropic magnetoresistance (AMR) of a polycrystalline ferromagnetic metal layer. In this case $\rho^\mathrm{AMR}_\mathrm{long}$ is described by the well-known expression 
\begin{eqnarray}
   \rho^\mathrm{AMR}_\mathrm{long}&=&\rho^\mathrm{AMR}_{0}+ \rho^\mathrm{AMR}_{1} m_{j}^2  
	\label{eq:AMR}
	\; ,
\end{eqnarray}  
where $\rho^\mathrm{AMR}_{0}$ is given by the resistivity perpendicular to $\mathbf{m}$ ($\rho_{\bot}$) and $\rho^\mathrm{AMR}_{1}$ by the difference of the resistivities parallel ($\rho_{\parallel}$) and perpendicular ($\rho_{\bot}$) to $\mathbf{m}$ as $\rho^\mathrm{AMR}_{1}=\rho_{\parallel}-\rho_{\bot}$.\cite{McGuire:1975} We therefore expect an angle dependence of $\rho^\mathrm{AMR}_\mathrm{long}$, when rotating the magnetic field in the ip- and oopt-plane.  

The above equations are only valid for polycrystalline heavy-metal layers as it is the case for the standard Pt/YIG//YAG bilayer samples. For crystalline materials, the symmetry of the crystal has to be taken into account for the calculation of the resistivity tensor $\boldsymbol{\rho}$.\cite{Birss:1966,Limmer:2006,Philippi-Kobs:2019} For the inverted YIG/Pt//YAG bilayer sample types, we found that the Pt layer is weakly textured along the [111]-direction (see Fig.~\ref{fig:fig1}(b)). However, since the FWHM of the rocking curve along the Pt~(111) reflection is more than $4^\circ$, we neglect any contribution from the crystal symmetry to $\boldsymbol{\rho}$ also in the inverted Pt/YIG//YAG bilayer sample type in the following discussion.
%%
%--------------------------- Fig. 7 --------------------------
\begin{figure} [t]
  \includegraphics[width=\columnwidth]{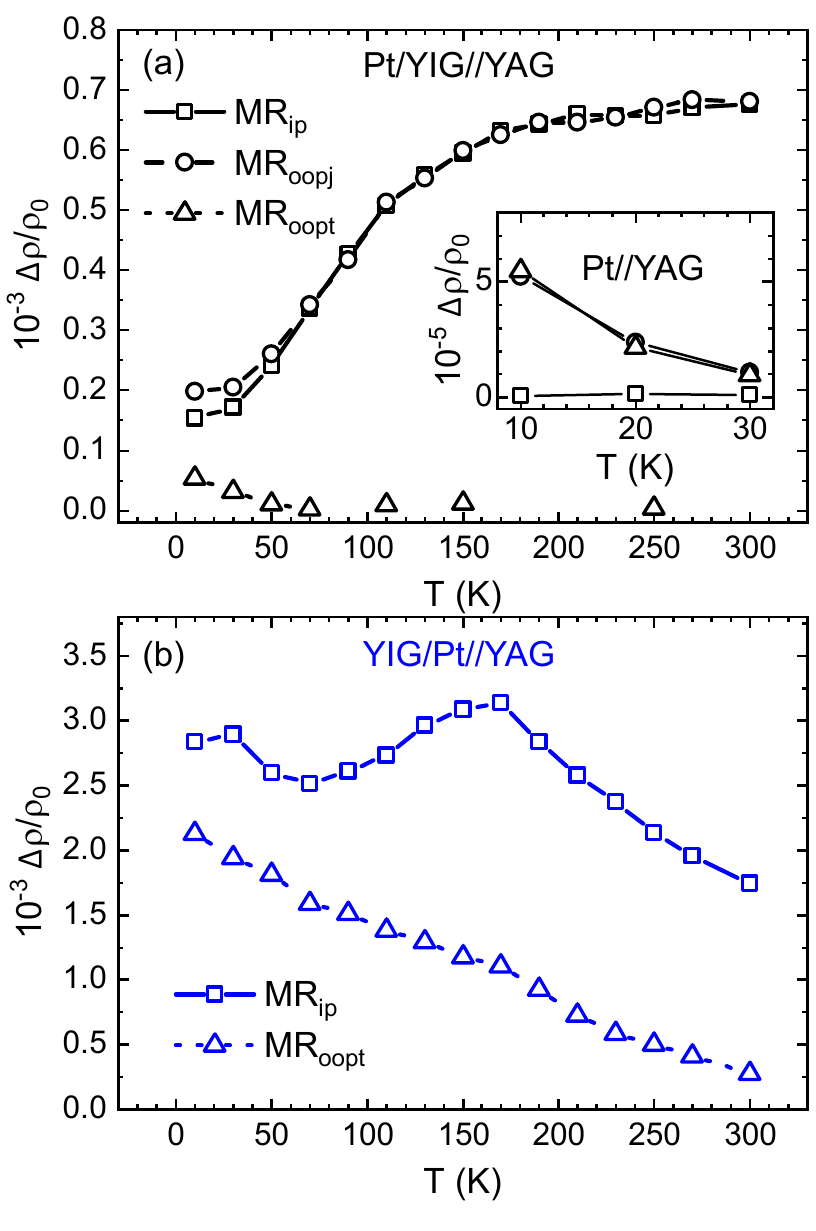}
  \caption{MR amplitudes obtained from ADMR measurements of (a) a standard Pt/YIG//YAG bilayer sample (black symbols) and (b) an inverted YIG/Pt//YAG bilayer sample (blue symbols) carried out at $\mu_0H=7$\,T in different rotation planes (ip, oopj, oopt) of the magnetic field. The inset shows the MR amplitudes of a Pt//YAG reference sample.}
 \label{fig:fig7}
\end{figure}
%--------------------------- Fig. 7 --------------------------
%%    

As shown in Figs.~\ref{fig:fig6}(a)-(c), the standard Pt/YIG//YAG bilayer sample displays a $\cos^2\alpha$ and $\cos^2\beta$ angle-dependence of $\rho_\mathrm{long}$ at all temperatures when rotating the magnetic field in the ip- and oopj-plane (cf.~Figs.~\ref{fig:fig6}(a),(b)). However, while no ADMR signal above the noise level of $4 \times 10^{-6}$ is visible for 250\,K and 110\,K in the oopt rotation plane, a small but finite ADMR is observed at 10\,K. We fit our data according to Eq.~(\ref{eq:SMR}) using $\cos^2$ functions (cf.~solid lines in Fig.~\ref{fig:fig6}(a)-(c)). The obtained SMR amplitudes $\Delta\rho/\rho_0$ from the ip-, oopj-, and oopt-measurements are depicted in Fig.~\ref{fig:fig7}(a). We observe a decrease of the SMR amplitudes with decreasing temperature $T$, which is in agreement with our previous report.\cite{Meyer:2014} However, since the SMR depends on the spin diffusion length, the spin Hall angle as well as the spin mixing conductance,\cite{Chen:2013} diverse reports of the $T$-dependence of the SMR can be found in literature due to different $T$-dependencies of these physical quantities in different Pt/YIG samples. In particular, the $T$-dependence of the SMR differs significantly for \textit{in-situ}\cite{Meyer:2014} and \textit{ex-situ}\cite{Marmion:2014} fabricated Pt/YIG bilayers as well as for samples with Ar$^+$-ion cleaning\cite{Niimi:2013,Velez:2016a} or chemical etching\cite{Shiomi:2014} of the YIG layer prior to the Pt deposition. This demonstrates that both intrinsic and extrinsic (phonon and impurity) scattering play an important role for the SMR.\cite{Isasa:2015,Karnad:2018} The $T$-dependence of the SMR can thus be regarded as a hallmark of the quality of Pt/YIG samples.

As obvious from Fig.~\ref{fig:fig6}(c) and Fig.~\ref{fig:fig7}(a), we observe a finite ADMR for temperatures $T<50$\,K, while rotating the magnetic field in the oopt-plane. This behavior can be attributed to a MR in our Pt thin films, since we also found a finite ADMR of similar magnitude at low temperatures in Pt//YAG reference samples (see inset of Fig.~\ref{fig:fig7}(a)). This MR results from an increase of the resistance with magnetic fields along the normal of the thin film ($\mathbf{n}$-direction), and is most likely caused by weak antilocalization effects.\cite{Niimi:2013,Velez:2016,Miao:2017} Since no MR is observed in ip-ADMR measurements of the Pt//YAG reference sample (square open symbols in the inset of Fig.~\ref{fig:fig7}(a)), the finite MR in oopt-ADMR measurements of the Pt/YIG//YAG sample cannot be related to a conventional AMR or MPE.\cite{Amamou:2018} It causes a difference of the MR amplitude recorded in ip- and oopj-ADMR measurements with $\mathrm{MR}_\mathrm{ip} < \mathrm{MR}_\mathrm{oopj}$ for $T < 50$\,K.

%%
%--------------------------- Fig. 8 --------------------------
\begin{figure}[t]
  \includegraphics[width=\columnwidth]{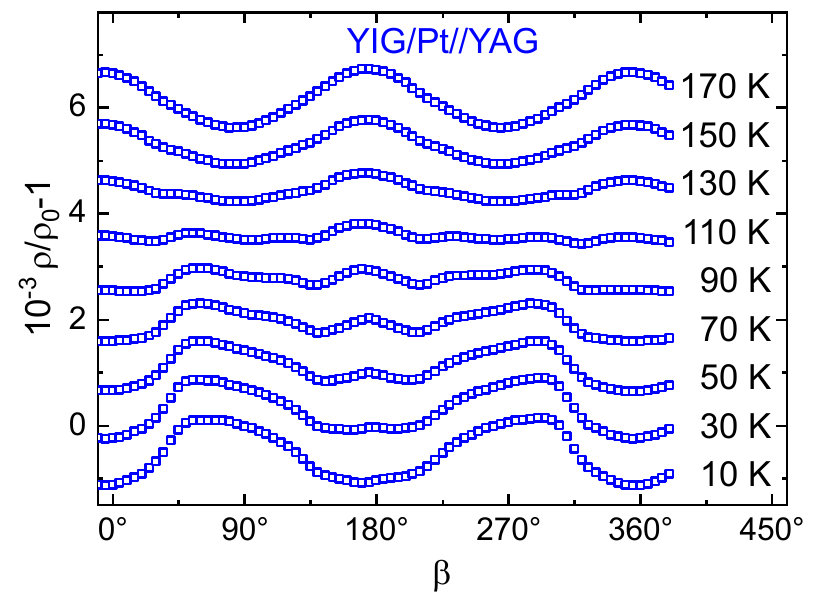}
  \caption{Normalized longitudinal resistivity $\rho_\mathrm{long}$ for the inverted YIG/Pt//YAG bilayer sample recorded while rotating the magnetic field in the oopj-rotation plane at different temperatures. The curves are vertically shifted for clarity.}
 \label{fig:fig8}
\end{figure}
%--------------------------- Fig. 8 --------------------------
%%

The inverted YIG/Pt//YAG bilayer sample, however, shows a qualitatively different angle-dependence of $\rho_\mathrm{long}$ (cf.~Fig.~\ref{fig:fig6}(d)-(f)). While $\rho_\mathrm{long}$ also follows a $\cos^2\alpha$-dependence in ip-rotations of the magnetic field (cf.~Fig.~\ref{fig:fig6}(d)), a $90^\circ$-phase shift is observed as a function of temperature in ADMR measurements, when rotating the magnetic field in the oopj-plane (cf.~Fig.~\ref{fig:fig6}(e)). Furthermore, a pronounced angle-dependence is found, when rotating the magnetic field in the oopt-plane (cf.~Fig.~\ref{fig:fig6}(f)). To analyze the data, we again use a $\cos^2$-function to fit the angle-dependence of $\rho_\mathrm{long}$ (cf.~solid lines in Fig.~\ref{fig:fig6}(d)-(f)). The obtained amplitudes $\Delta\rho$ normalized to $\rho_0$ of the ip- and oopt-measurements are shown in Fig.~\ref{fig:fig7}(b). 

By comparing the obtained MR values of the standard and the inverted bilayers (cf.~Fig.~\ref{fig:fig7}(a),(b)), we can attribute the angle-dependence of $\rho_\mathrm{long}$ in oopt-magnetic field rotations to a finite AMR contribution based on the finite Pt-induced magnetic moment in the inverted bilayer (see previous section). As expected the obtained AMR amplitude continuously increases with decreasing temperature. The magnetoresistance amplitude obtained from ip-measurements, however, can be explained by a superposition of a finite SMR effect, which dominates at high temperatures, and the AMR contribution prevailing at low temperatures. The cross-over from the AMR to the SMR behavior when increasing the temperature is also visible in the ADMR recorded during rotation of the magnetic field in the oopj-plane (see~Fig.~\ref{fig:fig8}). At $T=10$\,K, an angle modulation of $\rho_\mathrm{long}$ with maxima around $90^\circ$ and $270^\circ$ (i.e. for $\mathbf{H} \parallel \pm \mathbf{j}$) is observed. The deviations from the expected $\cos \beta^2$-dependence are most likely caused by magnetic anisotropy effects. By increasing the temperature, maxima at $\mathbf{H} \parallel \pm \mathbf{n}$ (at $\beta = 0^\circ, 180^\circ, 360^\circ$) appear leading to a complex angle-dependence of $\rho_\mathrm{long}$. These additional maxima can be related to an SMR-like angle-dependence according to Eq.~(\ref{eq:SMR}). For $T>150$\,K, this contribution dominates and a single $\cos^2 \beta$-dependence is visible. Figure~\ref{fig:fig8} clearly demonstrates that the angle-dependence of $\rho_\mathrm{long}$ can not be simply described by Eqs.~(\ref{eq:SMR}) and (\ref{eq:AMR}). Instead, a more complex resistance network has to be taken into account which requires a more detailed knowledge of the local microstructure of the Pt thin film.

\section{Conclusion}

In summary, we showed that in heavy metal (HM)/ferromagnetic insulator (FMI) bilayer thin film samples consisting of the HM Pt and the FMI YIG the appearance of magnetic proximity effects in the HM Pt crucially depends on the quality of the Pt/YIG interface. On standard Pt/YIG bilayer samples with a clean and sharp interface, we do not observe any indication of an induced magnetic moment in the Pt layer in x-ray magnetic circular dichroism (XMCD)\cite{Geprags:2012} as well as x-ray resonant magnetic reflectivity experiments (XRMR). In these samples, the observed magnetoresistance can be explained solely within the spin Hall magnetoresistance theory. In contrast, in inverted YIG/Pt bilayer samples, a finite induced magnetic moment of up to $0.058\,\mu_\mathrm{B}$/Pt can be found by XMCD and XRMR at room temperature, which increases at lower temperatures. This finite moment, which often is attributed to a magnetic proximity effect at a perfect Pt/YIG interface, is shown to originate from a finite interdiffusion at this interface due to the deposition of YIG on Pt, associated with both an elevated temperature and high kinetic energy of the atoms/ions impinging on the surface during the PLD process. In those samples, the spin Hall magnetoresistance is superimposed by an induced magnetic moment based anisotropic magnetoresistance, which becomes dominant at low temperatures. This demonstrates that a combined temperature-dependent x-ray and magnetotransport study is essential to confirm or exclude any magnetic proximity effects in HM/FMI heterostructures.   

\begin{acknowledgments}
This work was supported by the European Synchrotron Radiation Facility (ESRF) via HE-3784, HC-1500, HC-1783, HC-2058, and HC-3268 as well as the Deutsche Forschungsgemeinschaft (DFG) via SPP 1538 (Projects No.~GO 944/4-1 and No. KU 3271/1-1). We acknowledge Diamond Light Source (Didcot, UK) for time on Beamline I16 under Proposal MT12772-1 and DESY (Hamburg, Germany), a member of the Helmholtz Association HGF, for the provision of experimental facilities. Parts of this research were carried out at beamline P09 at PETRA III. The authors thank Andreas Erb for the preparation of the polycrystalline PLD target, Thomas Brenninger for technical support, Matthias Althammer for illustrations, Florian Bertram, Olga Kuschel, Timo Oberbiermann, Jari Rodewald as well as Jan Krieft, Anastasiia Moskaltsova, Tobias Pohlmann for help during the beamtimes at DLS and DESY, respectively. Technical support at the beamlines has been given by Ben Moser and Gareth Nisbet (beamline I16, DLS) as well as David Reuther (beamline P09, DESY).
\end{acknowledgments}

\bibliography{manuscript-proximity-and-smr}

\providecommand{\noopsort}[1]{}\providecommand{\singleletter}[1]{#1}%

\end{document}